# The SPARC Toroidal Field Model Coil Program

Zachary S. Hartwig, Rui F. Vieira, Darby Dunn, Theodore Golfinopoulos, Brian LaBombard, Christopher Lammi, Phil Michael, Susan Agabian, David Arsenault, Raheem Barnett, Mike Barry, Larry Bartoszek, William K. Beck, David Bellofatto, Daniel Brunner, William Burke, Jason Burrows, William Byford, Charles Cauley, Sarah Chamberlain, David Chavarria, JL Cheng, James Chicarello, Van Diep, Eric Dombrowski, Jeffrey Doody, Raouf Doos, Brian Eberlin, Jose Estrada, Vincent Fry, Matthew Fulton, Sarah Garberg, Robert Granetz, Aliya Greenberg, Martin Greenwald, Samuel Heller, Amanda Hubbard, Ernest Ihloff, James Irby, Mark Iverson, Peter Jardin, Daniel Korsun, Sergey Kuznetsov, Stephen Lane-Walsh, Richard Landry, Richard Lations, Rick Leccacorvi, Matthew Levine, George Mackay, Kristen Metcalfe, Kevin Moazeni, John Mota, Theodore Mouratidis, Robert Mumgaard, JP Muncks, Richard Murray, Daniel Nash, Ben Nottingham, Colin O'Shea, Andrew Pfeiffer, Samuel Pierson, Clayton Purdy, Alexi Radovinsky, Dhananjay K. Ravikumar, Veronica Reyes, Nicolo Riva, Ron Rosati, Michael Rowell, Erica E. Salazar, Fernando Santoro, Akhdiyor Sattarov, Wayne Saunders, Patrick Schweiger, Shane Schweiger, Maise Shepard, Syun'ichi Shiraiwa, Maria Silveira, FT Snowman, Brandon Sorbom, Peter Stahle, Ken Stevens, Joshua Stillerman, Deepthi Tammana, Thomas Toland, David Tracey, Ronnie Turcotte, Kiran Uppalapati, Matthew Vernacchia, Christopher Vidal, Erik Voirin, Alex Warner, Amy Watterson, Dennis G. Whyte, Sidney Wilcox, Michael Wolf, Bruce Wood, Lihua Zhou, Alex Zhukovsky.

*Abstract*— **The SPARC Toroidal Field Model Coil (TFMC) Program was a three-year effort between 2018 and 2021 that developed novel Rare Earth Yttrium Barium Copper Oxide (REBCO) superconductor technologies and then successfully utilized these technologies to design, build, and test a first-in-class, high-field (~20 T), representative-scale (~3 m) superconducting toroidal field coil. With the principal objective of demonstrating mature, large-scale, REBCO magnets, the project was executed jointly by the MIT Plasma Science and Fusion Center (PSFC) and Commonwealth Fusion Systems (CFS) as a technology enabler of the high-field pathway to fusion energy and, in particular, as a risk retirement program for the TF magnet in the SPARC net-energy fusion tokamak. Weighing 10,058 kg (22,174 lb) and utilizing 270 km (168 mi) of REBCO, the TFMC is a no-insulation magnet comprising a winding pack of sixteen REBCO stack-in-plate style pancakes and twoij termination plates inside a Nitronic-50 structural case, which also serves as a pressure vessel for the cryogenic coolant flowing through channels in the winding pack. To execute the TFMC tests, a new magnet test facility was built and commissioned at the MIT PSFC. A centerpiece of the test facility is a pair of 50 kA LN2-cooled REBCO binary current leads and VIPER REBCO cable feeder system. A novel liquid-free cryocooler-based cryogenic system provided 20 K supercritical helium. The magnet is integrated with the feeder and helium circulation system inside a large 20 m³ vacuum cryostat, which contains internal LN2 radiation shields and access for facility and magnet instrumentation.**

**The TFMC achieved its programmatic goal of experimentally demonstrating a large-scale high-field**

This paper was submitted to arXiv on August 18 2023. The SPARC TFMC Program was a joint effort between the MIT Plasma Science and Fusion Center (PSFC) and Commonwealth Fusion Systems (CFS). The work was funded by CFS, and the parties have pursued patent protection relating to inventions. CFS has exclusive commercial rights to the technology for energy generation.

Corresponding author: Zachary S. Hartwig <hartwig@psfc.mit.edu>

Author affiliations based on associated institution at the time of contribution.

Zachary S. Hartwig, Rui F. Vieira, Theodore Golfinopolous, Brian LaBombard, Phil Michael, Susan Agabian, David Arsenault, William K. Beck, David Bellofatto, William Burke, Jason Burrows, William Byford, Charles Cauley, James Chicarello, Jeffrey Doody, Vincent Fry, Matthew Fulton, Robert Granetz, Martin Greenwald, Amanda Hubbard, Ernest Ihloff, James Irby, Mark Iverson, Daniel Korsun, , Stephen Lane-Walsh, Richard Landry, Richard Lations, Rick Leccacorvi, Richard Murray, Andrew Pfeiffer, Samuel Pierson, Alexi Radovinsky, DJ Ravikumar, Nicolo Riva, Ron Rosati, Michael Rowell, Erica E. Salazar, Fernando Santoro, Wayne Saunders, Shane Schweiger, Syun'ichi Shiraiwa, Maria Silveira, FT Snowman, Peter Stahle, Joshua Stillerman, Thomas Toland, David Tracey, Christopher Vidal, Amy Watterson, Dennis G. Whyte, Bruce Wood, Lihua Zhou, and Alex Zhukovsky are with the MIT Plasma Science and Fusion Center, Cambridge MA 02139 USA.

Darby Dunn, Christopher Lammi, Raheem Barnett, Michael Barry, Daniel Brunner, Sarah Chamberlain, David Chavarria, JL Cheng, Van Diep, Eric Dombrowski, Raouf Doos, Brian Eberlin, Sarah Garberg, Aliya Greenberg, Samuel Heller, Peter Jardin, Sergey Kuznetsov, Chris Lammi, Matthew Levine, Kristen Metcalfe, Robert Mumgaard, JP Muncks, Daniel Nash, Ben Nottingham, Colin O'Shea, Clayton Purdy, Veronica Reyes, Akhdiyor Sattarov, Patrick Schweiger, Maise Shepard, Brandon Sorbom, Ken Stevens, Deepthi Tammana, Ronnie Turcotte, Kiran Uppalapati, Matthew Vernacchia, Alex Warner, and Sidney Wilcox are with Commonwealth Fusion Systems, Devens MA 01434 USA.

Larry Bartoszek is with Bartoszek Engineering, Aurora IL 60506 USA.

Erik Voirin is with eViorin Engineering Consulting, Batavia IL 60510 USA.

Michael Wolf is with Karlsruhe Institute of Technology, 76131 Karlsruhe Germany.

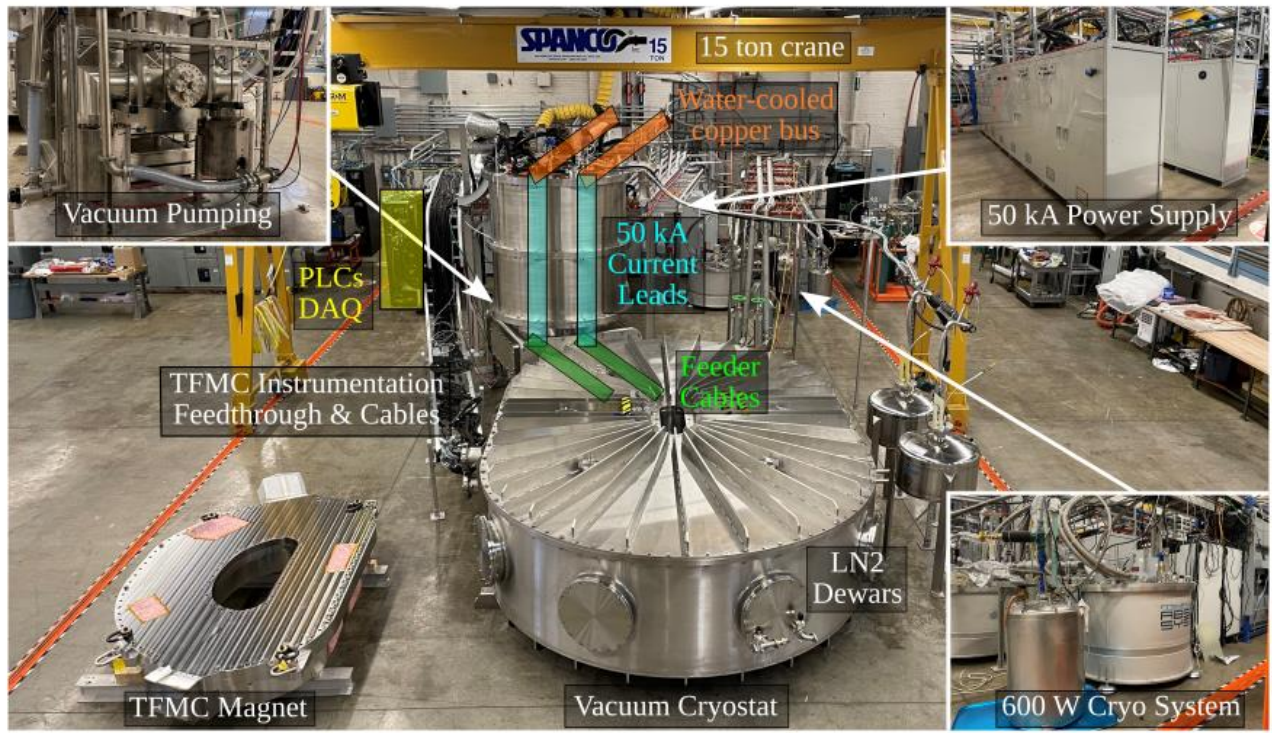

*Fig. 1: A view of the TFMC Magnet and test facility at the MIT Plasma Science and Fusion Center.*

REBCO magnet, achieving 20.1 T peak field-on-conductor with 40.5 kA of terminal current, 815 kN/m of Lorentz loading on the REBCO stacks, and almost 1 GPa of mechanical stress accommodated by the structural case. Fifteen internal demountable pancake-to-pancake joints operated in the 0.5 to 2.0 nΩ range at 20 K and in magnetic fields up to 12 T. The DC and AC electromagnetic performance of the magnet, predicted by new advances in high-fidelity computational models, was confirmed in two test campaigns while the massively parallel, single-pass, pressure-vessel style coolant scheme capable of large heat removal was validated. The REBCO current lead and feeder system was experimentally qualified up to 50 kA, and the crycooler based cryogenic system provided 600 W of cooling power at 20 K with mass flow rates up to 70 g/s at a maximum design pressure of 20 bar-a for the test campaigns. Finally, the feasibility of using passive, self-protection against a quench in a fusion-scale NI TF coil was experimentally assessed. While the TFMC was intentionally not optimized for quench resiliency – and suffered localized thermal damage in response to an intentional open-circuit quench at 31.5 kA terminal current – the extensive data and validated models that it produced represent a critical step towards this important objective.

*Index Terms*—Fusion energy, Rare Earth Barium Copper Oxide, Superconducting magnet, Toroidal field magnet

## I. INTRODUCTION

The SPARC Toroidal Field Model Coil (TFMC) Project was an approximately three-year effort between 2018 and 2021 to retire the design, fabrication, and operational risks inherent in large-scale, high-field superconducting magnets based on the high temperature superconductor (HTS) Rare-Earth Barium Copper Oxide (REBCO). Executed jointly between the MIT Plasma Science and Fusion Center (PSFC) and Commonwealth Fusion Systems (CFS), the project developed novel Rare Earth Yttrium Barium Copper Oxide (REBCO) superconductor technologies and then utilized those technologies to successfully design, build, and test a first-in-class, high-field (~20 T) representative scale (~3 m in linear size) superconducting toroidal field (TF) coil known as the TFMC. In parallel to the construction of the magnet, a new superconducting magnet test facility was established at MIT to serve as the proving grounds for the experimental demonstration of the magnet. The test facility itself incorporated novel advances in both REBCO and cryogenic technology to meet the technical and schedule requirements for testing the magnet. The magnet and test facility, shown together in Fig. 1, were combined in the fall of 2021 to carry out a series of experimental test campaigns to assess the performance of the magnet, validating the magnet modeling, design, and fabrication techniques and proving the novel technologies deployed in the test facility.

This paper, the first of six papers in this special issue covering the TFMC Program, serves two principal objectives: first, it presents a self-contained, high-level technical and programmatic overview of the entire TFMC Program, including a summary of the high-field path to fusion energy and a brief history of large-scale superconducting fusion magnet development programs; and second, it provides the context for understanding the accompanying five papers that cover specific technical areas of the TFMC Program. These papers focus on the following topics:

- Magnet design, fabrication, and assembly [1]
- Test facility design, construction, and commissioning [2]
- 50 kA binary LN2-cooled REBCO current leads [3]
- 600 W cryocooler-based helium cryogenic system [4]
- Results of the test campaigns and post-mortem [5]

Taken together, the papers attempt to provide a comprehensive review of the program's background, objectives, activities, and achievements.

This overview and introductory paper is structured as follows: Section II presents the context and motivation for the TFMC Program, namely the technical and programmatic advantages of accelerating the deployment of commercial fusion energy through the use of high-field superconducting magnets; Section III briefly reviews the history of large-scale superconducting magnet development in order to provide background and insight into why and how the TFMC Program was executed; Section IV provides a programmatic overview of the project as a foundation to understand the technical research and development that was conducted; Section V briefly reviews the two distinct conductor and coil technologies developed in the first phase of the TFMC Program (VIPER Cables [6] and no-insulation no-twist, or NINT, coils [7]) and summarizes the advantages of selecting NINT coils to be scale-up in the TFMC; Section VI summarizes the remaining three phases of the program, which are explored in greater technical depth in the accompanying five papers; and, finally, Section VII concludes with some summary remarks about the TFMC Program impact on the state of large-scale high-field REBCO magnets and the next steps towards high-field net-energy fusion devices.

## II. THE HIGH MAGNETIC FIELD PATH TO FUSION ENERGY

It has long been recognized that the strength of the primary magnetic field plays a principal role in determining the performance of a magnetic fusion energy device such as a tokamak. Since the first magnetic confinement devices in the 1950s, this has manifested in the exponential increase in plasma physics performance metrics achieved by a succession of increasingly high-field fusion tokamaks, which have relied on advances of the science and engineering of large-scale electromagnets. In fact, a straightforward theoretical analysis of the nuclear and engineering aspects of a tokamak, combined with a basic knowledge plasma physics constraints, shows that increasing the magnetic field strength is one of the most powerful and accessible means of achieving both the conditions required to access a stable, burning, net-energy plasma and the design of a cost-effective compact tokamak fusion power plant [8]. The optimization strategy of maximizing the magnetic field strength to achieve net-fusion energy in more compact, lower cost, and faster-to-build tokamaks has historically been known as the "high-field path to fusion energy," and it can be divided chronologically into two separate eras based on the magnet technology utilized.

The first iteration of the high-field path focused on the use of resistive copper magnet technology. The copper high-field path to fusion energy relied on advanced water- or LN2-cooled Bitter plate style magnets to achieve pulsed magnetic fields in tokamaks exceeding 12 T peak field-on-coil, corresponding to approximately 8 T in the plasma center [9]. Several tokamaks were built and operated that experimentally validated the plasma physics advantages of high-field operation and advanced the state of high magnetic field engineering including most notably the Frascati Tokamak Upgrade at ENEA in Italy [10] and the three Alcator tokamaks at MIT in the United States culminating in the Alcator C-Mod Tokamak [11].

Based on both the physics and engineering successes of FTU and the Alcators, high-field copper-based tokamaks, particularly in the United States, dominated the roadmap for achieving net-fusion energy in compact machines in the 1980s and 1990s. Several major machine design and engineering activities were initiated, including the Burning Plasma Experiment (BPX) [12]Compact Ignition Tokamak (CIT) [13], the Fusion Ignition Research Experiment (FIRE) [14], and the Ignitor Tokamak [15]. Despite the advantages ascribed to these machines, the copper high-field path was ultimately abandoned due to the challenge of scaling resistive copper magnet technology to fusion power plants and the preference to begin utilizing superconducting magnets, based on the emergent low-temperature superconductors (LTS) NbTi in the 1980s and then $Nb_3Sn$ in the 1990s, as an alternative.

Starting in 1978 with the T-7 tokamak in the Soviet Union, a series of superconducting machines with LTS were built and operated with moderate field ($1 - 8$ T in the plasma center). This list includes EAST in China [16], KSTAR in South Korea [17], SST-1 in India [18], T-7 and T-15 in the Soviet Union/Russia, Tore Supra / WEST in France [19], and TRIAM-1M in Japan [20] with JT-60SA [21] as the most recent. These machines have provided the necessary superconducting magnet engineering basis for ITER, a tokamak designed to achieve 500 MW of fusion power with a gain factor of ten using $Nb_3Sn$ TF coils to provide ~5.3 T in the center of the plasma [22]. Based on the design and initial engineering work for ITER, a series of post-ITER conceptual, such as CFETR in China [23], and demonstration power plants such as DEMO in the EU [24], are being planned. Because $Nb_3Sn$ magnets limit the magnetic fields in the plasma to approximately ~5.5 T, such tokamaks typically have major radii between 6 and 9 m to achieve sufficient plasma performance, leading to extraordinarily large devices associated with high capital cost, multi-decadal schedules, scale challenges in supply chain and assembly, and complex organizational issues.

In response to these challenges, and due to the emergence of a new class of superconductors capable of achieving much higher magnetic fields than LTS materials, a second iteration of the high-field path to fusion energy was proposed by MIT in the

mid-2010s [25]. Anchored in the record-setting plasma physics performance of Alcator C-Mod in 2016 [26], this new high-field path proposes the use of superconducting TF magnets exceeding 20 T peak field-on-coil (8 – 12 T in the plasma center) based on REBCO.

REBCO was discovered in 1987 [27]. While little was then known about its superconducting performance and an engineering-relevant form factor remained decades in the future, several theoretical papers emerged in the following few years that examined the advantageous physics, cost, and complexity implications of superconducting toroidal field coils approaching 20 T [24 -25]. Only relatively recently has sufficiently high performance REBCO coated conductor tape been characterized and manufactured at the industrial volumes and cost levels required to actually design and build large-scale high-field fusion magnets [30].

An important engineering study of a toroidal field (TF) magnet with modern REBCO superconductor was carried out in 2011 [31] as part of the conceptual design activities for the VULCAN tokamak [32]. The study was the first to highlight significant advantages of high-field REBCO-based magnets for tokamaks, including feasible TF magnets approaching 20 T peak field-on-coil, an optimized operational temperature of 20 K, tolerance of large nuclear heating, and demountable low-resistance joints. Since that time, several physics and engineering assessments reinforcing and extending the scale, schedule, cost, and plasma-physics advantages of high-field fusion energy powerplants based on REBCO magnets have been carried out for conventional aspect ratio tokamaks [33], spherical tokamaks [34], and stellarators [35].

In the mid-2010s, it was evident that execution of the high field path required accelerating the development and deployment of fusion-relevant REBCO conductors and coils beyond their then-nascent state. Two separate potential technologies had emerged by this time. First, a wide variety of high-current REBCO cables based on the insulated cable-in-conduit conductor (CICC) concept were being built and tested on small-scales. Although none to-date had demonstrated the required performance, robustness, and scalability required for fusion-scale magnets, the initial prototypes were promising [36]. Second, single REBCO tape wound no-insulation coils had demonstrated high field performance with simpler fabrication and the potential for self-protection during quench[37]; this represented an alternative, albeit very different, coil concept compared to insulated CICC cables.

Regardless of which REBCO-based concept would be developed as a basis for high-field superconducting fusion magnet technology, the TFMC Program engineering R&D would follow a two-part sequence. The first part would focus on technology readiness of basic conductor/coil technology at the small scale. Efforts would focus on engineering design, fabrication processes, computational modeling, and rigorous experimental testing of small-scale components. The overarching requirement emplaced on the processes and technology created during this phase was that it must be inherently scalable on rapid timescales. The second part would undertake the scale-up of the conductor/coil technology into a representative scale magnet, or "model coil", that would ultimately qualify the design, fabrication, modeling, and operation of a high-field REBCO magnet for readiness in a first-generation high-field net fusion energy device. Such model coils have historically played a critical role in advancing superconducting fusion magnets, confirming the arrival of a step-change in this key enabling technology.

## IV. BRIEF REVIEW OF LARGE-SCALE SUPERCONDUCTING FUSION MAGNET TEST PROGRAMS

Superconducting magnet systems for fusion embody substantial scale, cost, schedule, complexity, and risk. From a physics perspective, the design, operation, and ultimately fusion performance of the machine is fundamentally set by the magnetic fields, making achievement of the magnet design specifications imperative. From an engineering perspective, the magnet systems are heavily integrated into the device superstructure, making them difficult or impossible to replace or repair. Thus, while advances in superconducting magnet technology can often bring transformational benefits to a fusion machine, it is imperative that the technology be derisked before being utilized in a fusion device. Therefore, the integration of major advances in superconducting magnet technology has historically been preceded by signficant research and development programs on specially designed "model coils" that achieve fusion device magnet relevant scale and performance but often in a stand-alone configuration in specialized test facilities for efficiency.

The first such endeavor was the Large Coil Task (LCT), an international collaboration between the United States, Japan, Switzerland, and Europe that sought to evaluate the feasibility of large-scale superconducting magnets for fusion tokamaks [38]. By 1987, the LCT successfully demonstrated multiple fusion-relevant superconducting magnet technologies, many for the first time in a large superconducting coil. Six 2.5 m x 3.5 m bore toroidal field (TF) like coils capable of producing 8 T peak field-on-coil in steady-state were built and tested individually and as an array at the International Fusion Superconducting Magnet Test Facility (IFSMTF) at Oak Ridge National Laboratory (ORNL) [39]. Five coils – General Dynamics (GD), General Electric/ORNL (GE/ORNL), Switzerland (CH), EUROATOM (EU) and Japan (JA) – utilized NbTi superconductor in a steel structure while a sixth from Westinghouse (WH) was the first large-scale coil to utilize react-and-wind $Nb_3Sn$ in an aluminum structure. To evaluate cryogenics, three coils (GD, GE/ORNL, JA) were cooled with atmospheric liquid helium pool boiling while the remaining coils (EU, CH, WH) employed forced-flow supercritical helium at 1.5 MPa. To evaluate winding configurations, five of the coils were pancake wound while the GD coil was layer wound.

The LCT proved that steady-state, or DC, superconducting coils could indeed be built and operated at fusion-relevant scales and performance; however, it was recognized that the pulsed, or AC, central solenoid (CS) and poloidal field (PF) magnets contained significant further challenges and would require their own model coil programs. Two such programs were undertaken. In the late 1980s and early 1990s, the US and Japan built and tested several Demonstration Poloidal Coils (DPC), including a 2 m bore 30 kA, 10 T/s Nb₃Sn cable-in-conduit conductor (CICC) coil led by MIT in the US [40] and 30 kA NbTi coil led by JAERI in Japan [41]. Following the conclusion of the DPC project, the large POLO model coil, a PF-like NbTi coil with a 3 m bore, 15 kA nominal current, and 2 T/s ramp rate was built and successfully tested by Forschungzentrum Karlsruhe (FZK) in 1997 [42].

One of the key conclusions of ITER Conceptual Design Activity (CDA) was that the ITER magnet systems would need to be superconducting; thus, the follow-on ITER Engineering Design Activity (EDA) initiated two model coil programs in 1992 with the explicit objective of demonstrating the design, manufacturing, and operation of ITER-relevant TF and CS magnets. The ITER Toroidal Field Model Coil (TFMC), carried out by the European Union, built a 40 ton, 80 kA, 7.8 MA-turn Nb₃Sn CICC coil using 7 double pancakes inside a SS316 LN structural case with the objective of maximizing likeness to the ITER TF coil in a scaled test article [43]. The coil was successfully tested, individually and mounted at a 4.5 degree angle to the EURATOM LCT coil to increase peak field-on-coil and out-of-plane IxB Lorentz loading to 9.97 T and 797 kN/m, respectively, at the TOSKA facility at FZK in 2001 [44]. Concurrently, the ITER Central Solenoid Model Coil (CSMC) project, executed by the European Union, Japan, Russia, and the United States, built a 101 ton, 46 kA coil composed of an outer and inner module, built by Japan and the US, respectively, along with an insert coil built by Japan. Tested at the JAERI facilities in Naka, Japan, the combined CSMC achieved a peak field-on-coil of 13 T with ramp rates of 0.6 T/s (inner module) and 1.2 T/s (insert coil) – in excess of ITER CS requirements at the time – with a total stored energy of 640 MJ with no performance degradation after 10,000 load cycles [45].

Following a similar progression as ITER, the MIT PSFC and CFS concluded an approximately five-year conceptual design activity that resulted in the physics basis for the SPARC tokamak in 2020 [46]. One of the key conclusions was that a tokamak approximately ~35 times smaller in volume than ITER could achieve high fusion gain ($Q_{physics} > 2$) provided that (1) the TF magnet could provide 12.2 T on-axis magnetic field, corresponding to approximately 22 T peak field-on-coil and (2) that the TF magnet could maintain cryostability despite the significant nuclear heating expected in a compact tokamak with minimal radiation shielding. Both requirements rule out the use of NbTi or Nb₃Sn superconductor, which can practically achieve maximum fields of around 9 T and 13 T, respectively, in fusion-style magnets and cannot tolerate significant nuclear heating due to their small margins to critical temperature, low heat capacity, and low thermal diffusivity at 4 K. The only superconductor capable of meeting these requirements was REBCO.

*Table I: Parameter comparison between the TFMC and one coil from the SPARC toroidal field (TF) magnet in 2021.*

| Design Parameter | TFMC | TF Coil |
|---|---|---|
| Magnet mass [kg] | 10,058 | 18,025 |
| Magnet size [m] | 1.9 x 2.9 | 3.0 x 4.3 |
| Winding pack (WP) mass [kg] | 5,113 | 7,975 |
| WP minimum turn radius [m] | 0.2 | 0.4 |
| WP current density [A/mm2] | 153 | 94 |
| WP inductance [H] | 0.14 | 0.59 |
| WP amp-turns [MA-turns] | 10.4 | 6.3 |
| Terminal current [kA] | 40.5 kA | 31.3 |
| Number of turns | 256 | 200 |
| Number of pancakes | 16 | 16 |
| Total REBCO [km] | 270 | 270 |
| Coolant type | Supercritical helium | |
| Coolant pressure [bar] | 10 – 20 | 15 |
| Operating temperature [K] | 20 | 8 – 17 |
| Peak magnetic field [T] | 20.1 | 23 |
| Peak Lorentz loading [kN/m] | 822 | 750 |
| Magnetic stored energy [MJ] | 110 | 316 |

Given the lack of experience world-wide with the design, fabrication, and operation of fusion-scale REBCO magnets, MIT and CFS undertook a rapid one-year conductor and coil development program to establish the foundational magnet technologies followed by a two-year period to design, build, and test the TFMC within a new magnet test facility.

IV. PROGRAMMATIC OVERVIEW OF THE TFMC PROGRAM

A. Project Objectives

The principal function of the TFMC was as a risk-retirement article for the SPARC TF magnet; as such, the technical requirements and the project objectives flowed down from the TF to the TFMC. A comparison between the defining magnet engineering parameters of the TFMC and the SPARC TF magnet are presented in Table I. Compared to the SPARC TF in terms of size, the TFMC is an approximately 55% scaled version, which is large enough to develop representative fabrication processing but at significantly reduced cost and schedule compared to TF-scale. Despite the scaled size, the TFMC is capable of matching or exceeding many of the critical parameters including the peak field-on-coil, electromagnetic loading on conductor, terminal electrical current, winding pack current density, and cryogenic cooling metrics. Although the stored magnetic energy does not match, the TFMC at 110 MJ provides sufficient energy to assess the feasibility of dissipation uniformly in the winding pack in a quench scenario and to induce localized thermal damage if this cannot be achieved.

With the technical requirements defined, the TFMC Program was then programmatically oriented around the achievement of six main objectives:

1.  The achievement of 20 T peak field-on-conductor in a large-bore magnet with a terminal current of 40 kA and a transverse IxB load of 800 kN/m at a temperature of 20 K with total stored magnetic energy of 110 MJ.

2.  The demonstration of key aspects of the magnet design:

    - Confirmation of no-insulation magnet physics and operation, particularly the current, voltage, power, and temperature distributions in the magnet during charging/discharging and steady-state regimes.

    - Demonstration of efficient cryogenic cooling through the use of massively parallel, single pass, machined cooling channels within the winding pack and the structural case as a pressure vessel.

    - Demonstration of simple, robust, demountable pancake-to-pancake joints internal to the magnet winding pack with resistances in the 1 nΩ range.

3.  The development and validation of high-fidelity electromagnetic and thermo-mechanical models of no-insulation magnets such that confidence in the design and operation of the full-scale SPARC TF could be achieved.

4.  Exploration of passive, self-protecting resilience to current sharing and quench in representative-scale, fusion-relevant no insulation REBCO magnets.

5.  The development and qualification of materials, instrumentation, tooling, fabrication processes, and external vendors to enable confidence and speed in the construction of the SPARC TF magnet.

6.  Development of the REBCO supply chain by providing magnet-pull in the form of challenging tape specification, close technical partnerships, and, most importantly, an unprecedented injection of capital expenditure through large tape orders to enable manufacturer scale-up.

### B. Project Constraints

Another key shaping function for the TFMC, its test facility, and the test campaigns were the constraints imposed upon the project. These constraints were of two types: anticipated and accepted as the boundary conditions of executing the project under the circumstances available; and unanticipated from external sources beyond project control. Understanding the constraints helps provide insight into why certain technical decisions were made and how those decisions manifested. While a comprehensive list is beyond the scope of this paper, a brief review of the key constraints and mitigating decisions is included here.

Budget and schedule are two important constraints in any large-scale engineering project but were amplified in the TFMC Program as a result of project sponsorship through a start-up company (CFS) with a fixed capital raise and scheduled milestone targets. Schedule proved to be the most defining constraint of the two. The June 2021 completion date for the TFMC Program was set at the official start of the MIT-CFS colloboration in June of 2018, providing an immutable 3-year window to complete the four phases of the project described in Section IV C.

Several strategies were employed to minimize schedule. The "make-buy" strategy was carefully evaluated with heavy weighting towards in-housing major parts and subsystems at MIT and CFS while choosing fabrication processes and partner vendors primarily on the ability to execute successfully on rapid schedules over other considerations such as cost. Closely coupled to this strategy was the high – but carefully managed – tolerance for technical risk in the R&D process, allowing signficant advances with less iterations. Examples of these strategies on the TFMC include the decision to build and then use all the equipment required to wind the magnet in-house while intentionally crafting the size and shape of the TFMC to fit within the existing capacities of established vendors who could deliver on the required schedule. Another example: external, existing test facilities were sought but none identified that could be ready on the schedule required, resulting in a decision to build a new test facility from scratch at MIT PSFC. Decisions for major elements of the test facility itself, such as the decision to design and build custom 50 kA binary LN2-cooled REBCO current leads, were driven by the lack of rapidly available 50 kA current leads from national laboratory or commercial vendors.

Another important constraint that shaped the project, in particular the design of the test facility, was the pre-existing infrastructure available at the MIT PSFC. Unlike almost all large-scale magnet test facilities, the MIT PSFC does not host an extensive liquid helium infrastructure, which is typically used both to provide the cryogenic coolant to the magnet as well as the coolant to superconducting current leads and feeder duct systems. To provide cooling to the TFMC, an innovative liquid-free cryocooler-based helium circulation system was developed with an external vendor to provide 600 W at 20 K cooling capability at significantly reduced cost, footprint, and schedule compared to a new liquid helium infrastructure. To provide cooling to the 50 kA superconducting current leads and feeder cables, the large 18,000 gallon capacity LN2 storage and distribution system was utilized by designing custom LN2-cooled REBCO current leads rather than the more traditional helium vapor-cooled leads found in similar scale magnet systems [47].

While high bay experimental halls were available for the fabrication and assembly of the TFMC and the 50 kA current leads, the test facility had to be built in an experimental hall with 18 foot vertical clearance. This constraint challenged the design of the vertical 50 kA current leads as well as resulted in the decision to test the TFMC magnet in the horizontal configuration, which departs from previous large-scale superconducting magnet tests such as the ITER TFMC in the KIT TOSKA facility [44].

A final important and completely unanticipated set of constraints were those imposed on the project by the COVID-19 pandemic. Perhaps the most significant impact was near-complete cessation of onsite hardware activities at MIT and CFS from mid-March to mid-May of 2020, followed by a slow ramp-up during the summer months of 2020. Many of the project's key vendors were similarly impacted, resulting in closures and staffing issues. Further challenges resulted from the inability of TFMC personal to routinely make onsite visits to vendors to collaborate effectively, witness critical processes, and perform quality inspections, typically a critical set of activities for successful complex engineering projects. Strategies employed to mitigate impacts included rotating two and three onsite shifts to minimize personnel density, requiring full personnel protection equipment and eventually vaccination while working on site, twice a week onsite COVID testing at MIT and CFS, moving all meetings to videoconferencing, and remote video inspection of vendor parts.

## C. TFMC Program Structure and Timeline

To achieve the programmatic objectives within the imposed project constraints as described in the previous two subsections, the TFMC Program was factored into four distinct phases spanning approximately three years:

1. *Foundational conductor/coil development (2018-2019):* Novel REBCO conductor and coil technologies were required to meet the requirements of superconducting fusion magnets capable of achieving in excess of 20 T peak field-on-coil. Activities in this phase focused on small-scale prototyping and testing of two very different base technologies suitable for such magnets: REBCO cables based on the insulated cable-in-conduit conductor (CICC) concept pioneered by Montgomery, Hoenig, and Steeves at MIT in the mid-1970s [48]; and REBCO coils based on the single tape-wound no-insulation coils developed by Hahn, Park, Bascunan, and Iwasa at MIT in the late 2000s [49].

2. *Magnet (2019-2021):* After demonstration of the base conductor/coil technologies, it was decided that a large-scale toroidal field (TF) magnet of fusion-relevant size and performance should be designed, fabricated, and tested. This approach followed a well-established precedent in the advancement of large-scale superconductor technology, as reviewed in Section IV, of building representative "model coils" as an important risk-retirement step towards constructing full-scale fusion devices incorporating that technology. Magnet fabrication and assembly took place at the MIT PSFC, in a large ~370 m² (~4,000 sq. ft.) hall explicitly reconfigured for this purpose with REBCO quality-assurance/quality-control and some magnet winding activities taking place at CFS.

3. *Magnet Test Facility (2019-2021):* In parallel with the magnet activities, the magnet required the establishment of a new test facility capable of meeting both the unique technical and demanding schedule requirements imposed on the project. The test facility was built at the MIT PSFC in order to repurpose the pre-existing large-scale experimental facilities and infrastructure made available by the shuttering of the Alcator C-Mod tokamak in 2016. These capabilities included a large ~835 m² (~9,000 sq. ft.) experimental hall, over 1 MVA of available electrical power up to 13.8 kV, a 400 kW distilled chilled water system, and 18,000 gallon LN2 distribution capabilities. Importantly, decades of experience conducting large-scale complex electromechanical experiments, combined with close cooperation with MIT's Environmental Health and Safety office, maximized the probability of safe, successful tests.

4. *Experimental Test Campaigns (2021):* Two distinct test campaigns were carried out in the fall of 2021. The first test campaign in August and September 2021 targeted assessment of the charging/discharging and steady-state performance of the coil including the demonstration of 20 T peak field-on-conductor, electromagnetic characteristics, measurement of joint resistances, and structural assessment of the coil under peak loads. The second test campaign in October 2021 focused on characterizing the response of the coil to off-normal events, including in current-sharing regimes at increasingly higher temperatures and under worst-case open circuit quench conditions. The test campaigns were followed by a series of non-destructive and destructive post-mortem analysis of the magnet to further validate the engineering design of the coil as well as to confirm experimental findings from the test campaigns, in particular the localized damage sustained during the final programmed open circuit quench.

## V. PRELIMINARY R&D TOWARDS THE TFMC

This section provides a brief overview of the development of two different REBCO technologies –VIPER cable and NINT coils – that took place during the first phase of the TFMC Program. This development was completed rapidly and in parallel at small-scale over approximately one year for two purposes: first, to determine which of the two technologies would be selected for the DC TF magnet in SPARC; and second, to develop a foundation for the cables required for the AC magnets in SPARC and superconducting feeder/bus systems. NINT, due primarily to the long L/R charging/discharging time constant imposed by NI magnet physics, is only suitable for the steady-state TF magnet. In contrast, insulated VIPER cable can satisfy the requirements of both steady-state magnets like the TF as well as handle the rapid magnetic flux density swings required by pulsed magnets like the central solenoid (CS) and poloidal field (PF) coils in a tokamak.

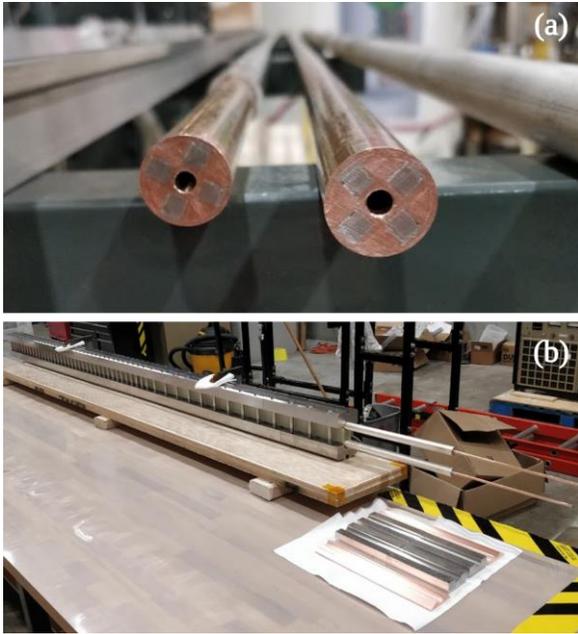

Fig. 2: VIPER cables. (a) shows a cross section of two of steel-jacketed VIPER cables configured for SULTAN testing. (b) shows a completed test assembly at MIT before shipping to PSI.

## A. The VIPER REBCO Cable

At the start of the TFMC Program, a range of fusion-relevant REBCO cable designs had been proposed [50]–[55] by institutions around the world. These cables, were all based on either the twisted stacked tape conductor (TSTC) architecture developed by Takayasu at MIT PSFC [56] or the conductor on round core (CORC®) architecture developed by van der Laan at Advanced Conductor Technologies [57]. Despite significant progress, several technical issues remained unresolved including critical current ($I_c$) degradation under representative high-field magnet electromagnetic loading (including axial strain and cycling) and quench detection given the slow normal zone propagation velocities inherent in HTS materials. Complexity in fabrication, particularly in the preparation of low resistance joints, and little development in integrating and testing these cables in magnet geometries presented scale-up challenges.

A REBCO CICC cable R&D program was undertaken to directly address these remaining issues. The result was the VIPER cable shown in Fig. 2, an industrially scalable high current REBCO cable based on the TSTC architecture. In the period of 2018-2019, over a dozen VIPER cables were built spanning lengths from 1 to 12 m, including a 3D single-turn coil intended to be tested in the NIFS 13 T large-bore test facility [58] and a multiturn pancake coil tested in LN2 at the MIT PSFC. The most stringent tests were a series of four experimental campaigns (comprising two identical VIPER cables each) at the SULTAN test facility at PSI [59]. Novel cable assemblies were developed to provide simultaneous transverse IxB Lorentz loads and axial strain to emulate the force state of a 3D coil [60]. Key results included achieve stable

$I_c$ with less than 5% degradation at IxB load of 382 kN/m (2000 cycles) and with axial strain of 0.3% for over 500 cycles [6], robust demountable joints in the 2-5 n$\Omega$ range [6], and the first demonstration of two separate fiber optic quench detection technologies on full-scale conductors in fusion-relevant conditions suitable for the low normal zone propagation velocity of REBCO [61].

## B. No Insulation – No Twist (NINT) REBCO Coils

In the mid-2010s, no-insulation REBCO (NI) technology had emerged as an alternative means of building high-field superconducting magnets. Superconducting NI coils were first proposed by Berlincourt and Hake, the co-discovers of NbTi as the one of the earliest practical superconductors, in the mid-1960s [62]. Since that time, and with the incorporation of REBCO as the superconductor, single-tape wound REBCO NI coils have developed into mature, high-field, well-engineered magnet systems and presently hold the world record of 45.5 T magnetic field strength [63]. The first fusion-relevant REBCO NI coil at the MIT PSFC was built in 2016 using 500 m of 12 mm wide tape and achieved a peak field in the 8 cm clear bore of 6 T [64].

Several challenges existed at the start of the TFMC Program in adapting such coils for fusion purposes. First, the scale of fusion TF coils required handling several GJ of stored magnet energy in the event of a quench, orders of magnitude beyond what had been experimentally demonstrated to date on small-bore NI coils. Second, the extremely long L/R charging/discharging times caused by high inductance from the single-tape/many-turns approach was incompatible with the timescales of fusion TF systems. Finally, the tightly packed NI coil geometry was unfavourable to efficiently removing the large amount of nuclear heat that is found in tokamaks, especially compact machines with limited space for nuclear radiation shielding.

In parallel with the VIPER cable program, a NINT REBCO coil R&D program was started to assess the feasibility of adapting NI coils to serve as fusion TF magnets. The first part of the program focused on conceptual coil design and resulted in several innovations. A REBCO stack-in-plate concept was developed in which a stack of REBCO tapes was wound directly into spiral grooves machined into one side of a plate of structural metal [7]. As was done for VIPER cables, REBCO tape stacks are soldered in place. Channels machined into the other side could accommodate cryogenic gas and be used for efficient, local, single-pass cooling [53]. After this, a two-pronged approach was taken that coupled a series of experiments on small demonstration coils with an advanced modelling program. A series of three 40 cm radius NINT coils with 16 tape REBCO stacks were built and tested at self-field at 77.3 K in LN2 and up to 10 kA terminal current at 15 K in a conduction-cooled configuration with the thermally insulated coil suspended over a bath of liquid helium as the coolant source, as shown in Fig. 3. These tests provided experimental data on the electromagnetic behaviour of the coils, particularly on the high cryostability of the coils and passive self-protection during quench by uniformly dissipating the magnet stored energy throughout the cold mass of the magnet.

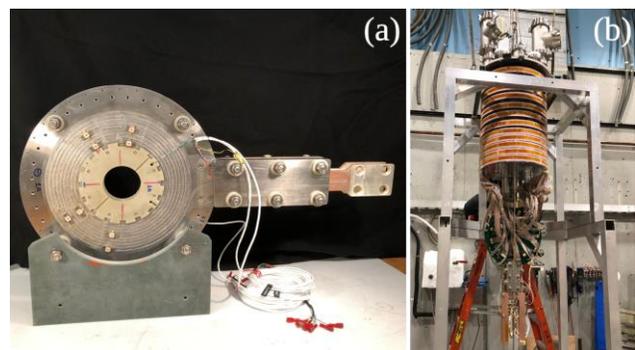

Fig. 3: NINT coils. (a) shows a NINT coil including the superconducting terminals and instrumentation cabling. (b) shows a NINT coil configured for 15 K testing at the MIT PSFC.

*Table II: Summary of the proposed NINT advantages for a tokamak TF coil used to select NINT technology for the TFMC winding pack.*

| Proposed NINT Design Features | NINT Advantage to be Assessed by TMFC |
|---|---|
| High winding pack current density | Compact, high-field REBCO magnet; Large magnet design space |
| High thermal stability | Resistant to quench; Robust to REBCO defects and local damage |
| Quench resiliency | Potential to eliminate active quench detection or mitigation systems |
| Single-pass pressure vessel cooling | Handle high nuclear heating; Local cooling optimization; Simple manifolding |
| Simple modular construction | Rapid fabrication; Scalable for SPARC and commercial use; Maintenance options |
| Intrinsically low voltage ($< 1$ V) | Minimal insulation; Simple fabrication; Low voltage leads and feeders; Safety |

*C. Selection of NINT for the TFMC winding pack*

At the conclusion of the first phase of TFMC in June 2019, a review was held to assess the results from the VIPER and NINT development programs with the objective of selecting the technology that would be scaled-up into the TFMC and, if successful, the SPARC TF. Design and analysis based on the experimental and modelling results from the VIPER and NINT development programs showed that either technology could successfully be used as the base magnet winding pack technology to achieve the necessary requirements for both the TFMC and the SPARC TF. Ultimately, NINT technology was selected for the TFMC on the basis of the proposed advantages summarized in Table II although, as discussed at the end of this section, VIPER cable was utilized in several ways in the TFMC Program.

The advantages can be roughly grouped into three types: superconducting; cryogenics; and fabrication. The high winding pack current density achievable with NINT, primarily due to the lack of insulation and the compact untwisted REBCO stack, provides a large design space for a compact but high-field magnet. Thermal stability is enabled by efficient cooling and large current sharing capacity of the unit cell. This allows the use of REBCO with defects such as dropouts, aggressively grading the REBCO tape stack, and possible tolerance to some threshold-level of damage sustained during fabrication or operation. Small-scale NI coils have demonstrated a significant degree of self-protection from quenches in experiments [64] and modelling [65]. If this capability could be successfully extended to large-scale NI coils suitable for fusion, active quench detection and mitigation systems could be eliminated. This would reduce the complexity of magnet fabrication and operation and provide passive protection, eliminating or minimizing perhaps the most significant operational challenge in superconducting fusion magnets.

In terms of cryogenics, the NINT design enables a scheme of massively parallel, single-pass cooling channels in the winding pack that maximizes global heat removal and can be optimized for local heat removal. This capability is important for the TF magnet in the SPARC tokamak, which minimizes nuclear radiation shielding to achieve compact size and results in large nuclear heating of the cryogenic TF.

In terms of fabrication, the open geometry of the NINT cable-in-plate concept, straightforward fabrication processes, and the absence of high voltage insulation makes production of NI magnets relatively efficient and scalable. In particular, the use of REBCO as opposed to LTS materials and the absence of an epoxy vacuum pressure impregnation step eliminates two of the most complex, long duration high temperature heat treatment operations involved in traditional insulated LTS CICC superconducting fusion magnets. Because the magnet winding pack is not encased in VPI epoxy and because demountable pancake-to-pancake joints can be used, NINT magnets can be fully disassembled for maintenance or component replacement without destructive operations. Finally, the intrinsic low voltage nature of NINT eliminates the need for high voltage current leads and feeder cables, traditionally one of the most challenging parts of insulated superconducting fusion magnet systems, and provides enhanced personnel and machine safety.

The VIPER cable R&D program was also drawn upon for TFMC. The successful vacuum pressure impregnation solder process developed for VIPER cables was directly adapted to fabricate NINT pancakes for the TFMC winding pack [66]. The robust, demountable, low resistance joints of the VIPER cable were modified to serve as demountable pancake-to-pancake joints embedded within the winding pack. Finally, VIPER cables were used directly in the implementation of the test facility, with three pairs of jointed VIPER cables forming the superconducting feeder system between the 50 kA LN2-cooled binary current leads and the TFMC magnet terminals.

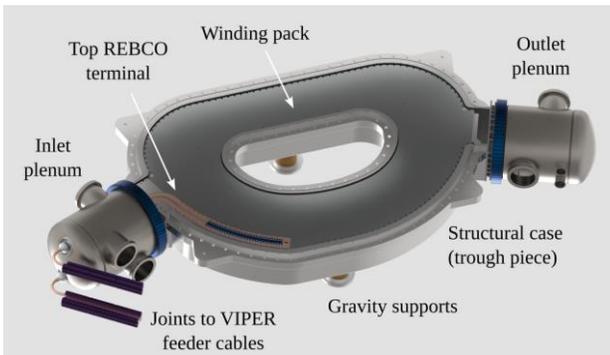

## VI. TFMC Program technical overview

The purpose of this section is to provide a short summary of the five technical papers that accompany this overview paper, covering major facets of the program at a high-level and providing a unified view of how the entire project was integrated together to form a complete whole. Readers interested in obtaining an in-depth technical analysis are encouraged to review the companion papers associated with each topic in this special issue.

### A. The TFMC Magnet

The TFMC, shown at the bottom-left of Fig. 1 and as a detailed CAD rendering in Fig. 4, is a non-insulated, REBCO stack-in-plate, style superconducting magnet. It weighed 10,058 kg and utilized 270 km of REBCO. It has three main components: (1) the winding pack; (2) the structural case; and (3) the plena. The winding pack is composed of a stack of 16 single pancakes with 2 termination plates top and bottom. The pancakes are Nitronic-40 radial plates machined with spiral channels on one side for the REBCO tape stack and single-pass channels on the opposite side for supercritical helium coolant. After REBCO winding and pancake assembly is complete, the pancake undergoes a vacuum-pressure impregnation solder process to provide good mechanical protection of the REBCO stack and efficient thermal and electrical connectivity throughout each pancake. Each pancake is electrically tested in LN2 following the solder process, providing quality assurance / quality control as well as superconducting performance data to guide pancake location within the winding pack stack and inform models of magnet performance. The pancakes are bolted at the perimeter to provide mechanical and thermal connectivity while inter-pancake joints provide low resistance current transfer between pancakes. The top and bottom termination plates facilitate electrical connection to a superconducting feeder system.

The winding pack is contained within a structural case, a "trough and lid" style design composed of two Nitronic-50 forgings machined to shape and bolted together. The case reacts the large electromechanical loads, with stresses approaching 1 GPa during operation, and serves as a pressure vessel to enable single-pass 20 bar supercritical helium flow that cools the winding pack and case. Two high pressure vessels or "plena" are attached to the case with unique high-pressure feedthroughs to provide winding pack access for current, cooling, and instrumentation, completing the magnet assembly.

### B. The TFMC Test Facility

The 835 m² (9,000 sq. ft.) TFMC Test Facility is shown in overview in Fig. 1. It was built as a stand-alone large-scale REBCO magnet test facility at the MIT PSFC in less than two years in the large experiment hall formerly housing power equipment for the Alcator C-Mod tokamak. The major engineering systems enabling magnet testing are as follows:

- *Main vacuum cryostat:* A 35-ton SS316 vacuum cryostat with 20 m³ internal volume was designed at MIT and built at an external vendor. External SS316 ribs minimize deflection under vacuum while the support structure underneath was carefully designed to properly support the approximately 10-ton TFMC magnet. The cryostat contains ten radial NW500 flanges to provide abundant internal access. A central bore is provided for instrumentation access within the high-field region of the TFMC. The top lid separates for installation/removal of large components.

- *Vacuum system:* Pumping is provided by two Leybold Turbovac TMP1000 (total of 2000 l/s pumping power) backed by Leybold Ecodry 35 scroll pumps. The scroll pumps were located outside of the 35-gauss line of the TFMC with the turbos enclosed within 1-inch thick iron magnetic shields. High vacuum in the $10^{-6}$ to $10^{-7}$ torr range was routinely achieved during experiments.

- *LN2 radiation shields and distribution system:* Radiation shields, consisting of modules of two steel panels with a thin interstitial space for LN2, were also designed at MIT and built by an external vendor. They were assembled within the vacuum cryostat. The LN2 was gravity fed from two storage dewars adjacent to the vacuum cryostat, which were provided with LN2 from an 18,000 external gallon storage tank.

- *Power supply:* The system was delivered by Alpha Scientific Electronics and is composed of eight cabinets each providing 6.25 kA for a total of 50 kA.

- *Current leads and feeder system:* A set of binary LN2-cooled 50 kA REBCO current leads and a VIPER cable feeder system were designed and built in-house at MIT PSFC. These systems are described in more detail in Section VI C.

- *Supercritical helium circulation system:* A liquid-free cryocooler-based system provided cooling down to 20 K for the TFMC, current leads, and feeder system was built

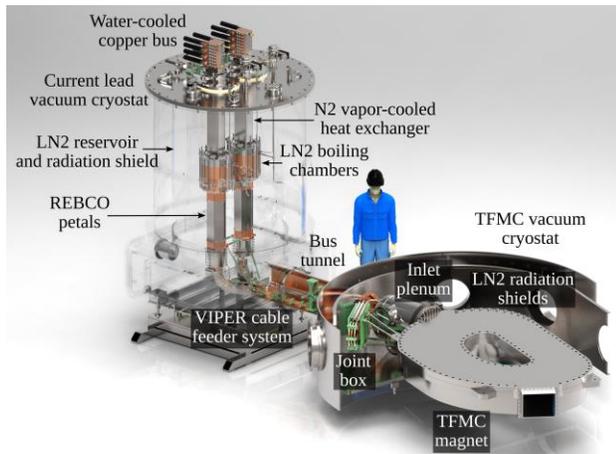

*Fig. 5: CAD rendering showing how the superconducting current leads (at top-left) and VIPER cable feeder system (at center) interface with the TFMC magnet (at bottom-right).*

by an external vendor. The system is described in more detail in Section VI D.

### C. Current leads and feeder cable system

The current leads (CL) and feeder cables (FC), shown together in proximity to the TFMC magnet in Fig. 5, represent a significant superconducting achievement in their own right. Designed and built in-house at the MIT PSFC as part of the TFMC Test Facility, the system was designed to handle 50 kA of current and utilize the significant onsite LN2 storage and distribution system available. Prior to the TFMC test campaigns, the CL and FC system was successfully commissioned at 41 kA to ensure the nominal 40.5 kA of terminal current for the TFMC tests; post-TFMC testing demonstrated stable operation at 50 kA, the maximum current available from the power supply. CL-FC joint resistances and FC joint resistances were measured in the 1 to 2 n$\Omega$ range at 40.5 kA.

Each of the approximately 3 m tall CLS consists of three sections. The upper, electrically resistive heat exchanger section is composed of a large heavily slotted C101 copper piece, designed to transport electrical current with high conductivity while being cooled with gaseous nitrogen boiling off from the middle section. The middle section is a large boiling chamber, containing an internal reservoir of LN2. Internally machined cooling fins maximize heat transfer while external claw pumps provide the ability to subcool the LN2 to 65 K by reducing the vacuum pressure in the chamber. The boiling chambers are fed from a large LN2 reservoir housed within the current lead vacuum cryostat, which also serves as the cryostat radiation shield. Connected to the bottom of the boiling chamber is the lower superconducting section. This section is composed of six individual petals that combine to provide an $I_c$(77 K, self-field) of 58.4 kA. Each petal consists of two large C101 copper terminal blocks spanned with a SS316 bridge containing twenty-two stacks of 4 REBCO tapes soldered into machined channels. When assembled the non-REBCO side of the C101 copper terminal blocks of the six petals forms an electrical and mechanical joint with the first VIPER feeder cable.

The superconducting FC system is composed of three sets of highly shaped non-planar VIPER cables, which are enclosed by a conduction cooled copper radiation shield attached to the main vacuum cryostat LN2 radiation shields. The longest, central set of cables – the "cold bus" – is a VIPER cable with a 10 mm central cooling hole for coolant. The coolant is warm supercritical helium exhaust from the TFMC, nominally at $22 - 25$ K compared to the 20 K helium inlet temperature to the magnet. Four stacks of REBCO resulted in an $I_c$(25 K, self-field) of 101 kA and a $T_c$(45 kA, self-field) of 55 K. This provides a large factor of safety to the 40.5 kA 25 K nominal operating conditions. The bus cables have a gradual "S" shape bend along the cable axis, which enables the feeder system to absorb with less than 0.1% axial strain the physical coefficient of thermal expansion movement induced during cooling when the current leads and the TFMC shrink away from each other. On either end of the bus cables, a set of nearly identical performance VIPER cables joins the bus to the CLs and the TFMC terminals; however, these cables have a solid copper former to maximize conduction cooling and simplify the supercritical helium circuit. All cables had to be bent to tolerances of a few mm to guarantee successful assembly.

### D. Cryocooler-based helium circulation system

Another innovative feature of the TFMC Test Facility was the implementation of a liquid-free cryocooler-based system that circulates supercritical helium as an alternative the more traditional but cost- and schedule-intensive liquid helium infrastructure found at large-scale magnet test facilities. The system was responsible for cooling the TFMC, the FCs, and the REBCO-section of the CLs.

The design and construction of the system was contracted to an external vendor. To meet the cryogenic requirements, two nearly identical modules – each containing four Cryomech AL630 cryocoolers housed within an LN2-shielded vacuum cryostat – were operated in parallel. Custom gas heat exchangers were integrated directly into the cryocooler coldhead, and cryofans with a maximum rotation speed of 60 krpm in each of the modules actuated helium circulation. During commissioning, the system confirmed 600 W of cooling power at 20 K with a supercritical helium mass flow of 70 g/s at maximum operating pressure of 20 bar-a. Heaters in each module provide for temperature control of the helium supply to the circuit.

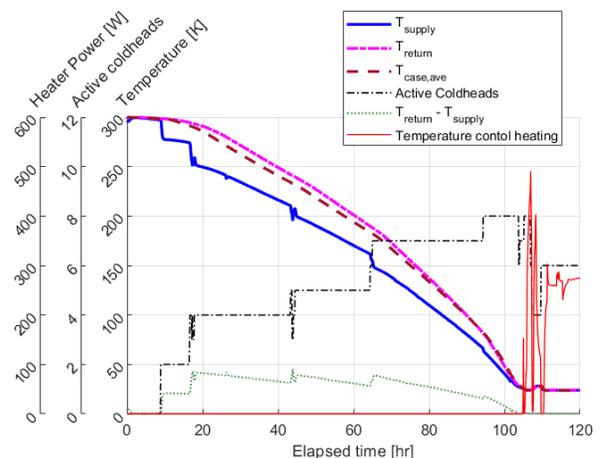

*Fig. 6: Data from the second cooldown of the TFMC magnet from 300 K to 20 K over the course of 5 days.*

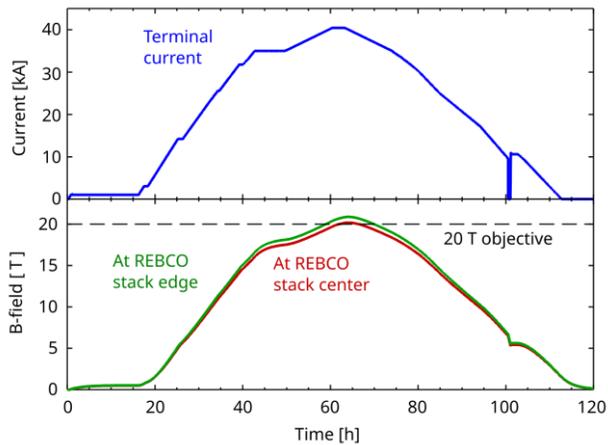

*Fig. 7: Data from the first TFMC test campaign, showing the terminal current (top) and 3D teslameter measurements of the peak field-on-conductor exceeding 20 T (bottom).*

Fig. 6 shows an example of cryogenic system performance during a 300 K to 24 K cooldown of the TFMC. Average helium circuit pressure, roughly maintained during cooldown by the addition of fresh helium, was approximately 13.3 bar-a. Cryocoolers were activated in sets of two to maximize the cooling rate but stay well within the administrative limit of a 50 K maximum difference between the helium inlet and return temperature to avoid temperature gradient-induced strains in the winding pack and structural case. The heaters are activated around hour 105 to maintain the target TFMC temperature of 24 K. The cooldown time of approximately 4.5 days was in good agreement with the in-house cryogenic circuit model used to design the helium circuit.

### E. Experimental results from the TFMC campaigns

The TFMC was experimentally assessed in two test campaigns in between August and October of 2021; following the test campaigns the coil was removed from the cryostat, carefully disassembled, and subject to a series of destructive and non-destructive tests. Disassembly was aided by the modular, demountable nature of the TFMC coil. The post-mortem was conducted to maximize learning from the TFMC, including confirmation of the engineering design and construction of the coil after rigorous testing up to full performance at 20 T and to assess the resiliency of the magnet to two open-circuit events that occurred during testing. The open-circuits were of particular interest, as they represent the most damaging circumstances for an NI coil, as currents driven radially cause internal joule heating that can lead to a quench.

Without the ability to detect-and-dump the stored magnetic energy as is the case for insulated magnets, large-scale NI fusion magnets must have a successfully strategy for handling quench. The TFMC tests provided the first opportunity to gather extensive data on a fusion-scale NI TF coil to aid in the validation of computational models and to guide future magnet design.

The first test campaign objectives were to assess the charging/discharging and steady-state electromagnetic response of the coil at the full design performance of 20 T peak field-on-conductor with 40 kA of terminal current. Measurements were successfully made of radial resistance, current, voltage and temperature distribution in the winding pack, the magnet's L/R decay time constant, internal pancake-to-pancake joint resistance, cryogenic helium flow distribution and cooling power, and structural performance. Fig. 7 shows an overview of the campaign's magnet ramp. The test took approximately 5 days due to the approximately 3-hour L/R decay time constant, placing great demands upon the operators. The magnet achieved a peak field-on-coil over a significant section of first turn REBCO stack of 20.1 T approximately 65 hours into the test.

The second test campaign had two objectives: to precisely quantify TFMC REBCO superconducting performance in DC as a function of temperature; and to fully measure the dynamics of the coil to an intentional open-circuit quench at a terminal current (31.5 kA) very close to the proposed SPARC TF magnet. Excellent voltage and temperature data were acquired during the initiation, incubation, and dump phases of the quench. The acquisition of this data set, and subsequent utilization to validate and extend the extensive array of NI computational modelling tools, achieved of one the most important objectives of the TFMC Program.

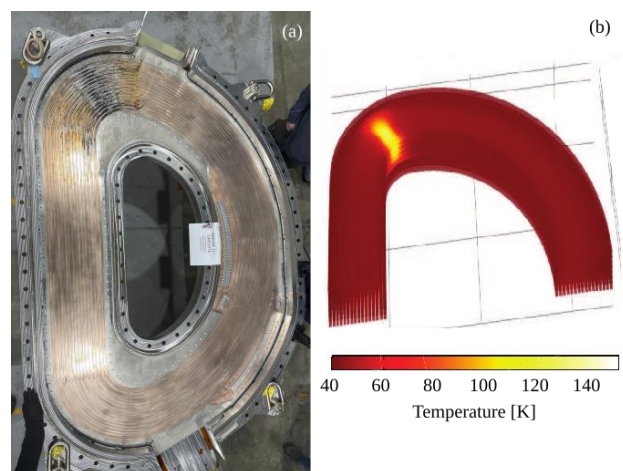

*Fig. 8: Post-quench analysis of the TFMC. (a) shows Pancake #12 during the post-mortem with a sharply defined region of thermal damage in the upper-left tight corner. (b) shows a 3D FEA simulation 170 s into the evolution of a 30 kA quench; the burn region is reproduced almost perfectly.*

In the quench; the predicted rapid (~3 s) inductive turn-to-turn and pancake-to-pancake quench cascades were clearly observed, confirming the dynamics of the basic self-protection mechanism for large-scale NI coils. The cascades are intended to rapidly distribute the stored magnetic energy uniformly throughout the magnet. Post-quench analysis and experiments, however, indicated the present of localized damage. Data on the global temperature distribution of the magnet during quench compared with 3D FEA model predictions indicated non-uniform energy deposition within the winding pack. Post-quench electromagnetic tests after recooling to 24 K found that the electromagnetic response of the coil (current path through the winding pack, azimuthal current and magnetic field, and total pancake voltages) had been altered through the upper half of the winding pack.

The TFMC's post-mortem confirmed the presence of localized damage within the upper half of the magnet winding pack. Concentrated in only a few pancakes with the epicentre in Pancake #12, thermal damage was found in a tightly defined azimuthal arc in one of the tight corners as shown in Fig. 8a. Due to the non-azimuthally symmetric shape, magnetic flux density is concentrated in the tight corners and creates a sharply defined critical surfaces characterized by high ratios of I/Ic (or T/Tc). For these regions, the result is sustained current sharing, leading to rapid temperature rise and ultimately burning in these areas, before the rapid inductive cascade can dissipate the magnetic stored energy throughout the magnet. The TFMC, which intentionally concentrated magnetic flux in the corners to achieve the programmatic DC performance goals, was inherently vulnerable to this effect. Indeed, the post-mortem confirmed that the engineering design, fabrication, and assembly of the coil were executed perfectly with no technical, operational, or other issue inducing the quench.

Importantly, this effect was predicted by several of the computational models developed in the TFMC program as shown in Fig. 8b; however, the models had not converged to a single self-consistent scenario requiring the experimental open circuit quench test. The extensive and unprecedented data set obtained from the experimental models have been used to refine the simulation toolset. The advances in NI magnet physics understanding and validated models are now being used to design next-generation NI TF coils that maximize quench resilience through a multipronged approach. These developments are outside the scope of this paper but expected to be published in the future.

## VII. Conclusion

The TFMC achieved its programmatic goal of demonstrating a large-scale high-field magnet, achieving 20.1 T peak field-on-conductor with 40.5 kA of terminal current, 815 kN/m of Lorentz loading on the REBCO stacks, and almost 1 GPa of mechanical stress. Internal demountable pancake-to-pancake joints operated in the 0.5 to 2.0 n$\Omega$ range at 20 K and in magnetic fields up to 12 T. The DC and AC electromagnetic performance of the magnet, predicted by new advances in high-fidelity computational models, was confirmed in experiment while a novel cryogenic coolant scheme capable of the large heat removal required by compact tokamaks was validated. A critical experimental step was taken to assess the feasibility of passive, self-protection against a quench in a fusion-scale NI TF coil. While the TFMC was intentionally not optimized for quench resiliency, the extensive data and validated models that it produced were required as an essential step towards this important objective.

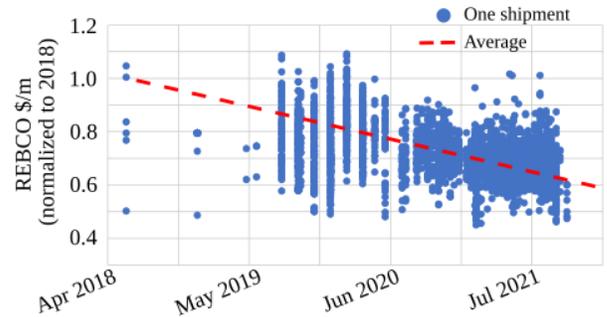

*Fig. 9: The decrease in normalized REBCO tape cost per meter during the time period spanning the TFMC Program..*

As a result of the TFMC program, design and fabrication of the NINT-based TF magnet, VIPER-based CS and PF magnets are underway, and new commercial magnets are underway Efforts since the close of the TFMC Program have focused on optimizing the TF towards quench resiliency, utilizing the improved modelling capabilities, technical innovations, and subsequent experiments. Based on the results from the TFMC Program, VIPER-based cables known as PIT-VIPER cable for the CS and PF are being developed and qualified in small-scale and model coil programs, making use of substantially expanded fabrication and test facilities at CFS as well as MIT PSFC. The TFMC Test facility, upgraded to support fast-ramp capabilities, is presently being utilized in cable and model coil tests for the CS. In addition to SPARC, CFS is also now designing and building high-field REBCO magnets for other scientific and industrial purposes, such as the WHAM high-field mirror project [67].

The TFMC Program also had an unprecedented impact on the REBCO tape industry. One of the objectives of the program was to reduce the cost of REBCO and catalyze the evolution of the industry from small, bespoke tape deliveries to standardized, industrial volumes. Fig. 9 shows the impact of the TFMC Program on REBCO cost per meter. Starting with initial MIT procurements from ten manufacturers in early 2018 through all of the CFS procurement for the TFMC in mid-2021, the TFMC was instrumental in reducing the average REBCO tape costs per meter by almost 40%. Procurements by CFS for the SPARC tokamak, now underway and expected to approach 10 million meters, should continue to decrease cost. Furthermore, the demanding REBCO specifications required by TFMC presented a technical challenge, some manufacturers, in response, proved capable of dramatically increasing both the performance and volume of REBCO at reduced cost [30].

Finally, the need for high volume $I_c(B,T,\theta)$ measurements as part of the NI modeling effort and quality-assurance / quality-control program for TFMC produced superior REBCO characterization equipment [68].

A final positive impact of the TFMC Program has been the significant acceleration of other high-field superconducting magnet efforts based on REBCO across a variety of fields. For example, the first non-planar stellarator coil from REBCO was designed, built, and tested with VIPER cable by MIT PSFC and Type One Energy, a private fusion company [69]. Another example is how the reduction in REBCO tape cost and demonstration of high-field REBCO magnets and ancillary technology is opening a path to next-generation collider experiments at the frontier of particle physics [70].

After a little over three years, the TFMC Program concluded in the fall of 2021 having successfully completed all of its programmatic and technical objectives. The state of REBCO magnet technology for fusion energy at the outset of the program, rooted in small-scale cable and coil prototypes and conceptual designs of full-scale system, looked very different at its close. The first representative-scale REBCO TF coil – in fact, the first large-scale REBCO magnet ever constructed – had been designed, built, and successfully tested. Ancillary technology, from high-current REBCO cables to 50 kA REBCO current leads, from liquid-free 600 W supercritical helium cryogenic systems to advanced 3D FEA modeling of the coupled thermal, mechanical, and electrical phenomena in large-scale NI magnets, had been successfully demonstrated. REBCO, in terms of performance, volume, and cost took its first significant step towards an industrial commodity product procured at the ton-scale. Finally, and perhaps most importantly, a new pathway to achieving fusion energy on accelerated timescales in compact devices, as well as opportunities in other areas of science and industry that utilize strong magnetic fields, had been opened by expanding the state of large-scale high-field superconducting magnets into the 20 T frontier.

## ACKNOWLEDGMENT

The authors are indebted to the many individuals and entities that made the TFMC Program possible. Executing such a program makes clear the full extent to which a tremendous number of people contribute to ensure collective success. We thank everyone who participated in the endeavour.

*At the MIT:* Professor Ron Ballinger for timely advice on metal forgings; Makoto Takayasu and Jim Kelsey for technical input and reviews; Joseph Minervini for leadership in superconductivity; Corinne Cotta (PSFC) for overseeing a sizeable fraction of TFMC procurement; Mary Davenport, Tesha Myers, Vick Sangha, and Katherine Ware for fiscal oversight; Jennifer James for providing unceasing administrative support; David Parker for handling PSFC operations; Karen Cote for handling PSFC safety; Brandon Savage, Lee Berkowitz, Mark London, and Clarence Tucker for tireless IT support; Ed Lamere and Mitch Galanek from MIT EHS for overseeing test program safety; Bob Armstrong and Randy Field at MIT Energy Initiative; MIT Central Machine Shop for fabrication support. MIT Central Utility Plant for keeping the water cold, and MIT facilities for the keeping the lights on.

*At CFS:* Andrea Jarrett and Joseph Stiebler who oversaw CFS procurement, Steve Renter for handling CFS operations, Carolina Zimmerman for collaboration on fiscal oversight, and Samuel Morgan for technical support.

*External:* Edward Moses for unwavering mentorship and project management guidance; Walter Fietz of Karlsruhe Institute of Technology for technical contributions and program review; Nick Strickland and Stuart Wimbush of Robinson Research Institute for collaboration on REBCO characterization; Nagato Yanagi of the National Institute for Fusion Sciences for collaboration on high-field cable testing; Satoshi Awaji, Tatsu Okada, and Arnaud Badel for opening their facilities and expertise to us in high-field REBCO $I_c(B,T,\theta)$ measurements; Jeff Mullins from Oliver Welding for perfect welds; to George Dodson (MIT), Joseph Minervini (MIT), Edward Moses, and Soren Prestemon (LBNL) for serving on the TFMC Safety and Operations Review committee.

*Vendors:* The project would not have succeeded without the commitment and execution of our vendor partners, who delivered complex procurements on challenging schedules while working at-risk through the COVID-19 pandemic. Our sincerest gratitude to each of you.

## REFERENCES

[1] R. Vieira, "The design, fabrication, and assembly of the SPARC Toroidal Field Model Coil," *IEEE Trans. Appl. Supercond.*, no. Special issue on the SPARC Toroidal Field Model Coil Project, 2023.

[2] T. Golfinopoulos, "The SPARC Toroidal Field Model Coil Magnet Test Facility," no. Special Issue on the SPARC Toroidal Field Model Coil, 2023.

[3] V. Fry, "50 kA capacity, nitrogen-cooled, demountable current leads for the SPARC Toroidal Field Model Coil," *IEEE Trans. Appl. Supercond.*, no. Special Issue on the SPARC Toroidal Field Model Coil Project, 2023.

[4] P. Michael, "A 20 K, 600 W, cryocooler-based supercritical helium system for the SPARC Toroidal Field Model Coil," *IEEE Trans. Appl. Supercond.*, no. Special Issue on the SPARC Toroidal Field Model Coil Project, 2023.

[5] D. Whyte, "Experimental results from the SPARC Toroidal Field Model Coil test campaigns.," *IEEE Trans. Appl. Supercond.*, no. Special Issue on the SPARC Toroidal Field Model Coil, 2023.

[6] Z. S. Hartwig *et al.*, "VIPER: an industrially scalable high-current high-temperature superconductor cable,"

*Supercond. Sci. Technol.*, vol. 33, no. 11, p. 11LT01, Oct. 2020, doi: 10.1088/1361-6668/abb8c0.

[7] B. LABOMBARD *et al.*, "Spiral-Grooved, Stacked-Plate Superconducting Magnets And Related Construction Techniques," US20200211744A1, Jul. 02, 2020 Accessed: Jul. 13, 2023. [Online]. Available: https://patents.google.com/patent/US20200211744A1/en

[8] J. P. Freidberg, F. J. Mangiarotti, and J. Minervini, "Designing a tokamak fusion reactor—How does plasma physics fit in?," *Phys. Plasmas*, vol. 22, no. 7, p. 070901, Jul. 2015, doi: 10.1063/1.4923266.

[9] D. R. Cohn and L. Bromberg, "Advantages of high-field tokamaks for fusion reactor development," *J. Fusion Energy*, vol. 5, no. 3, pp. 161–170, Sep. 1986, doi: 10.1007/BF01050610.

[10] R. Andreani, *The FTU Frascati Tokamak Upgrade*. United States: Pergamon Books Inc, 1987.

[11] M. Greenwald *et al.*, "20 years of research on the Alcator C-Mod tokamak)," *Phys. Plasmas*, vol. 21, no. 11, p. 110501, Nov. 2014, doi: 10.1063/1.4901920.

[12] G. Cargulia, C. Bushnell, and J. Citrolo, "Evolution of the BPX tokamak configuration," in *[Proceedings] The 14th IEEE/NPSS Symposium Fusion Engineering*, Sep. 1991, pp. 1047–1049 vol.2. doi: 10.1109/FUSION.1991.218767.

[13] D. Post *et al.*, "Physics Aspects of the Compact Ignition Tokamak," *Phys. Scr.*, vol. 1987, no. T16, p. 89, Jan. 1987, doi: 10.1088/0031-8949/1987/T16/011.

[14] D. M. Meade, "FIRE, a next step option for magnetic fusion," *Fusion Eng. Des.*, vol. 63–64, pp. 531–540, Dec. 2002, doi: 10.1016/S0920-3796(02)00282-X.

[15] B. Coppi and the Ignitor Project Group, "Highlights of the Ignitor experiment," *J. Fusion Energy*, vol. 13, no. 2, pp. 111–119, Sep. 1994, doi: 10.1007/BF02213946.

[16] S. Wu, "An overview of the EAST project," *Fusion Eng. Des.*, vol. 82, no. 5, pp. 463–471, Oct. 2007, doi: 10.1016/j.fusengdes.2007.03.012.

[17] G. S. Lee *et al.*, "Design and construction of the KSTAR tokamak," *Nucl. Fusion*, vol. 41, no. 10, p. 1515, Oct. 2001, doi: 10.1088/0029-5515/41/10/318.

[18] Y. C. Saxena and S.-1 Team, "Present status of the SST-1 project," *Nucl. Fusion*, vol. 40, no. 6, p. 1069, Jun. 2000, doi: 10.1088/0029-5515/40/6/305.

[19] R. Aymar, B. Bareyt, and G. Bon Mardion, "Tore Supra Basic design Tokamak system," France, 1980.

[20] H. Zushi *et al.*, "Overview of steady state tokamak plasma experiments in TRIAM-1M," *Nucl. Fusion*, vol. 43, no. 12, p. 1600, Dec. 2003, doi: 10.1088/0029-5515/43/12/006.

[21] Y. Kamada *et al.*, "Completion of JT-60SA construction and contribution to ITER," *Nucl. Fusion*, vol. 62, no. 4, p. 042002, Mar. 2022, doi: 10.1088/1741-4326/ac10e7.

[22] R. Aymar, P. Barabaschi, and Y. Shimomura, "The ITER design," *Plasma Phys. Control. Fusion*, vol. 44, no. 5, p. 519, Apr. 2002, doi: 10.1088/0741-3335/44/5/304.

[23] Y. T. Song *et al.*, "Concept Design of CFETR Tokamak Machine," *IEEE Trans. Plasma Sci.*, vol. 42, no. 3, pp. 503–509, Mar. 2014, doi: 10.1109/TPS.2014.2299277.

[24] G. Federici *et al.*, "Overview of EU DEMO design and R&D activities," *Fusion Eng. Des.*, vol. 89, no. 7, pp. 882–889, Oct. 2014, doi: 10.1016/j.fusengdes.2014.01.070.

[25] D. G. Whyte, J. Minervini, B. LaBombard, E. Marmar, L. Bromberg, and M. Greenwald, "Smaller & Sooner: Exploiting High Magnetic Fields from New Superconductors for a More Attractive Fusion Energy Development Path," *J. Fusion Energy*, vol. 35, no. 1, pp. 41–53, Feb. 2016, doi: 10.1007/s10894-015-0050-1.

[26] J. W. Hughes *et al.*, "Access to pedestal pressure relevant to burning plasmas on the high magnetic field tokamak Alcator C-Mod," *Nucl. Fusion*, vol. 58, no. 11, p. 112003, Sep. 2018, doi: 10.1088/1741-4326/aabc8a.

[27] J. G. Bednorz and K. A. Muller, "Possible highT c superconductivity in the Ba-La-Cu-O system," *Z. Phys. B Condens. Matter*, vol. 64, no. 2, pp. 189–193, Jun. 1986, doi: 10.1007/BF01303701.

[28] D. R. Cohn, J. Schwartz, L. Bromberg, and J. E. C. Williams, "Tokamak reactor concepts using high temperature, high-field superconductors," *J. Fusion Energy*, vol. 7, no. 1, pp. 91–94, Mar. 1988, doi: 10.1007/BF01108259.

[29] J. Schwartz, L. Bromberg, D. R. Cohn, and J. E. C. Williams, "A Commercial Tokamak Reactor Using Super High Field Superconducting Magnets," *Fusion Technol.*, vol. 15, no. 2P2B, pp. 957–964, Mar. 1989, doi: 10.13182/FST89-A39817.

[30] A. Molodyk *et al.*, "Development and large volume production of extremely high current density YBa2Cu3O7 superconducting wires for fusion," *Sci. Rep.*, vol. 11, no. 1, Art. no. 1, Jan. 2021, doi: 10.1038/s41598-021-81559-z.

[31] Z. S. Hartwig, C. B. Haakonsen, R. T. Mumgaard, and L. Bromberg, "An initial study of demountable high-temperature superconducting toroidal field magnets for the Vulcan tokamak conceptual design," *Fusion Eng. Des.*, vol. 87, no. 3, pp. 201–214, Mar. 2012, doi: 10.1016/j.fusengdes.2011.10.002.

[32] G. M. Olynyk *et al.*, "Vulcan: A steady-state tokamak for reactor-relevant plasma–material interaction science," *Fusion Eng. Des.*, vol. 87, no. 3, pp. 224–233, Mar. 2012, doi: 10.1016/j.fusengdes.2011.12.009.

[33] B. N. Sorbom *et al.*, "ARC: A compact, high-field, fusion nuclear science facility and demonstration power plant with demountable magnets," *Fusion Eng. Des.*, vol. 100, pp. 378–405, Nov. 2015, doi: 10.1016/j.fusengdes.2015.07.008.

[34] A. Sykes *et al.*, "Compact fusion energy based on the spherical tokamak," *Nucl. Fusion*, vol. 58, no. 1, p. 016039, Nov. 2017, doi: 10.1088/1741-4326/aa8c8d.

[35] J. A. Alonso *et al.*, "Physics design point of high-field stellarator reactors," *Nucl. Fusion*, vol. 62, no. 3, p. 036024, Mar. 2022, doi: 10.1088/1741-4326/ac49ac.

[36] P. Bruzzone *et al.*, "High temperature superconductors for fusion magnets," *Nucl. Fusion*, vol. 58, no. 10, p. 103001, Aug. 2018, doi: 10.1088/1741-4326/aad835.

[37] T. Lécrevisse, X. Chaud, P. Fazilleau, C. Genot, and J.-B. Song, "Metal-as-insulation HTS coils," *Supercond.*

*Sci. Technol.*, vol. 35, no. 7, p. 074004, May 2022, doi: 10.1088/1361-6668/ac49a5.

[38] D. S. Beard, W. Klose, and S. Shimamoto, "The IEA Large Coil Task," *Fusion Eng. Des.*, vol. 7, no. ½, pp. 3–230, 1988.

[39] M. S. Lubell *et al.*, "The IEA Large Coil Task Test Results in IFSMTF," *10. international conference on magnet technology (MT-10), Boston, MA, USA, 21 Sep 1987*, Jan. 01, 1987. https://digital.library.unt.edu/ark:/67531/metadc109384 3/ (accessed Jul. 13, 2022).

[40] M. M. Steeves *et al.*, "The US demonstration poloidal coil," *IEEE Trans. Magn.*, vol. 27, no. 2, pp. 2369–2372, Mar. 1991, doi: 10.1109/20.133694.

[41] K. Okuno *et al.*, "The first experiment of the 30 kA NbTi DEMO POLOIDAL COILS (DPC-U1 and -U2)," *Proc. MT-11*, vol. 2, pp. 812–817, 1989.

[42] M. Darweschsad *et al.*, "Development and test of the poloidal field prototype coil POLO at the Forschungszentrum Karlsruhe," *Fusion Eng. Des.*, vol. 36, no. 2, pp. 227–250, May 1997, doi: 10.1016/S0920-3796(97)00005-7.

[43] A. Ulbricht *et al.*, "The ITER toroidal field model coil project," *Fusion Eng. Des.*, vol. 73, no. 2, pp. 189–327, Oct. 2005, doi: 10.1016/j.fusengdes.2005.07.002.

[44] P. Komarek and E. Salpietro, "The test facility for the ITER TF model coil," *Fusion Eng. Des.*, vol. 41, no. 1, pp. 213–221, Sep. 1998, doi: 10.1016/S0920-3796(98)00116-1.

[45] H. Tsuji *et al.*, "Progress of the ITER central solenoid model coil programme," *Nucl. Fusion*, vol. 41, no. 5, pp. 645–651, May 2001, doi: 10.1088/0029-5515/41/5/319.

[46] A. J. Creely *et al.*, "Overview of the SPARC tokamak," *J. Plasma Phys.*, vol. 86, no. 5, Oct. 2020, doi: 10.1017/S0022377820001257.

[47] A. Ballarino *et al.*, "Design of the HTS Current Leads for ITER," *IEEE Trans. Appl. Supercond.*, vol. 22, no. 3, pp. 4800304–4800304, Jun. 2012, doi: 10.1109/TASC.2011.2175192.

[48] M. Hoenig and D. Montgomery, "Dense supercritical-helium cooled superconductors for large high field stabilized magnets," *IEEE Trans. Magn.*, vol. 11, no. 2, pp. 569–572, Mar. 1975, doi: 10.1109/TMAG.1975.1058601.

[49] S. Hahn, D. K. Park, J. Bascunan, and Y. Iwasa, "HTS Pancake Coils Without Turn-to-Turn Insulation," *IEEE Trans. Appl. Supercond.*, vol. 21, no. 3, pp. 1592–1595, Jun. 2011, doi: 10.1109/TASC.2010.2093492.

[50] G. Celentano *et al.*, "Design of an industrially feasible twisted-stack HTS cable-in-conduit conductor for fusion application," *IEEE Trans. Appl. Supercond.*, vol. 24, no. 3, 2014, doi: 10.1109/TASC.2013.2287910.

[51] T. Mulder, J. Weiss, D. van der Laan, A. Dudarev, and H. ten Kate, "Recent Progress in the Development of CORC Cable-In-Conduit Conductors," *IEEE Trans. Appl. Supercond.*, vol. 30, no. 4, pp. 1–5, Jun. 2020, doi: 10.1109/TASC.2020.2968251.

[52] N. Bykovsky, D. Uglietti, K. Sedlak, B. Stepanov, R. Wesche, and P. Bruzzone, "Performance evolution of 60 kA HTS cable prototypes in the EDIPO test facility," *Supercond. Sci. Technol.*, vol. 29, no. 8, p. 084002, Jun. 2016, doi: 10.1088/0953-2048/29/8/084002.

[53] N. Yanagi *et al.*, "Magnet design with 100-kA HTS STARS conductors for the helical fusion reactor," *Cryogenics*, vol. 80, pp. 243–249, Dec. 2016, doi: 10.1016/j.cryogenics.2016.06.011.

[54] M. J. Wolf, W. H. Fietz, C. M. Bayer, S. I. Schlachter, R. Heller, and K.-P. Weiss, "HTS CroCo: A Stacked HTS Conductor Optimized for High Currents and Long-Length Production," *IEEE Trans. Appl. Supercond.*, vol. 26, no. 2, pp. 19–24, Mar. 2016, doi: 10.1109/TASC.2016.2521323.

[55] M. Takayasu, L. Chiesa, P. D. Noyes, and J. V. Minervini, "Investigation of HTS Twisted-Stacked-Tape Cable (TSTC) Conductor for High-Field, High-Current Fusion Magnets," *IEEE Trans. Appl. Supercond.*, vol. 27, no. 4, pp. 1–5, Jun. 2017, doi: 10.1109/TASC.2017.2652328.

[56] M. Takayasu, L. Chiesa, L. Bromberg, and J. V. Minervini, "Cabling Method for High Current Conductors Made of HTS Tapes," *IEEE Trans. Appl. Supercond.*, vol. 21, no. 3, pp. 2340–2344, Jun. 2011, doi: 10.1109/TASC.2010.2094176.

[57] D. C. van der Laan, "YBa2Cu3O7-$\updelta$coated conductor cabling for low ac-loss and high-field magnet applications," *Supercond. Sci. Technol.*, vol. 22, no. 6, p. 065013, Apr. 2009, doi: 10.1088/0953-2048/22/6/065013.

[58] S. Imagawa *et al.*, "Plan for Testing High-Current Superconductors for Fusion Reactors with A 15T Test Facility," *Plasma Fusion Res.*, vol. 10, pp. 3405012–3405012, 2015, doi: 10.1585/pfr.10.3405012.

[59] R. Wesche, P. Bruzzone, D. Uglietti, N. Bykovsky, and M. Lewandowska, "Upgrade of SULTAN/EDIPO for HTS Cable Test," *Phys. Procedia*, vol. 67, pp. 762–767, Jan. 2015, doi: 10.1016/j.phpro.2015.06.129.

[60] V. Fry, J. Estrada, P. C. Michael, E. E. Salazar, R. F. Vieira, and Z. S. Hartwig, "Simultaneous transverse loading and axial strain for REBCO cable tests in the SULTAN facility," *Supercond. Sci. Technol.*, vol. 35, no. 7, p. 075007, May 2022, doi: 10.1088/1361-6668/ac6bcc.

[61] E. E. Salazar *et al.*, "Fiber optic quench detection for large-scale HTS magnets demonstrated on VIPER cable during high-fidelity testing at the SULTAN facility," *Supercond. Sci. Technol.*, vol. 34, no. 3, p. 035027, Feb. 2021, doi: 10.1088/1361-6668/abdba8.

[62] T. G. Berlincourt and R. R. Hake, "Means for insulating superconducting devices," US3187235A, Jun. 01, 1965 Accessed: Aug. 10, 2023. [Online]. Available: https://patents.google.com/patent/US3187235A/en?oq=US3187235A

[63] S. Hahn *et al.*, "45.5-tesla direct-current magnetic field generated with a high-temperature superconducting magnet," *Nature*, vol. 570, no. 7762, Art. no. 7762, Jun. 2019, doi: 10.1038/s41586-019-1293-1.

[64] P. C. Michael *et al.*, "Development of REBCO-Based Magnets for Plasma Physics Research," *IEEE Trans.*

*Appl. Supercond.*, vol. 27, no. 4, pp. 1–5, Jun. 2017, doi: 10.1109/TASC.2016.2626978.

[65] J.-B. Song, S. Hahn, T. Lécrevisse, J. Voccio, J. Bascuñán, and Y. Iwasa, "Over-current quench test and self-protecting behavior of a 7 T/78 mm multi-width no-insulation REBCO magnet at 4.2 K," *Supercond. Sci. Technol.*, vol. 28, no. 11, p. 114001, Sep. 2015, doi: 10.1088/0953-2048/28/11/114001.

[66] Y. Wang, W. K. Chan, and J. Schwartz, "Self-protection mechanisms in no-insulation (RE)Ba2Cu3Ox high temperature superconductor pancake coils," *Supercond. Sci. Technol.*, vol. 29, no. 4, p. 045007, Mar. 2016, doi: 10.1088/0953-2048/29/4/045007.

[67] A. Hubbard, "Processes, Systems and Devices for Metal-Filling of Open HTS Channels," 63/279,443

[68] J. Egedal, D. Endrizzi, C. B. Forest, and T. K. Fowler, "Fusion by beam ions in a low collisionality, high mirror ratio magnetic mirror," *Nucl. Fusion*, vol. 62, no. 12, p. 126053, Nov. 2022, doi: 10.1088/1741-4326/ac99ec.

[69] N. M. Strickland *et al.*, "Extended-Performance 'SuperCurrent' Cryogen-Free Transport Critical-Current Measurement System," *IEEE Trans. Appl. Supercond.*, vol. 31, no. 5, pp. 1–5, Aug. 2021, doi: 10.1109/TASC.2021.3060355.

[70] N. Riva, "Development of the first multi-turn non-planar REBCO stellarator coil using VIPER cable," *Supercond. Sci. Technol.*, no. Accepted for Publication, 2023.

[71] L. Bottura, S. Prestemon, L. Rossi, and A. V. Zlobin, "Superconducting magnets and technologies for future colliders," *Front. Phys.*, vol. 10, 2022, Accessed: Aug. 09, 2023. [Online]. Available: https://www.frontiersin.org/articles/10.3389/fphy.2022.935196